\documentclass[12pt]{article}

\usepackage{amsmath}
\usepackage{amssymb}

\usepackage{graphicx}

\usepackage[version=0.96]{pgf}
\usepackage{tikz}
\usetikzlibrary{arrows,shapes,snakes,automata,backgrounds,petri}
\usepackage[latin1]{inputenc}
\usepackage{verbatim}
\usepackage{cite}

\usepackage{color} 

\usepackage[labelfont=bf,labelsep=period,justification=raggedright]{caption}

\makeatletter
\renewcommand{\@biblabel}[1]{\quad#1.}
\makeatother

\date{}

\begin{document}

\begin{flushleft} 
{\Large{
\textbf{Statistical analysis of co-expression  properties of sets of genes in the mouse brain}}}\\
Pascal Grange, 
Partha P. Mitra 
\\
\bf Cold Spring Harbor Laboratory, One Bungtown Road, Cold Spring Harbor, New York 11724, USA
\\
$\ast$ E-mail: pascal.grange@polytechnique.org
\end{flushleft}

\section*{Abstract}

 We propose a quantitative method to estimate 
the  statistical properties of sets of genes for which expression data are available 
 and co-registered on a reference atlas of the brain. It is based on graph-theoretic
properties of co-expression coefficients between pairs of genes. We apply this method 
to mouse genes from the Allen Gene Expression Atlas. 
Co-expression patterns of a list of several hundreds of genes 
related to addiction are analyzed, using ISH data produced for the mouse brain 
at the Allen Institute. It appears that large subsets of this set of 
genes are much more highly co-expressed than expected by chance.

\tableofcontents

\section{Introduction}

In this era of complete genomes, genes are related to medical conditions 
by genome-wide association studies. This results in large lists  of condition-related genes.
It is desirable to prioritize some subsets of these lists for further study. 
 High-throughput experiments have already provided neuroscience with
a publicly-available dataset at a resultion of 200 microns for the mouse brain, the Allen  
 Gene Expression Atlas (AGEA), \cite{AllenGenome, images, neufoAllen, AllenAtlasMol}. The AGEA can be used as a reference set to 
 assess how exceptional a set of genes is. We define co-expression matrices 
for sets of genes, and study statistical properties of the underlying graphs by Monte Carlo methods.\\

These methods are applied to a set of 288 candidate genes extracted from the NicSNP database ,
{\ttfamily{http://zork.wustl.edu/nida/Results/data1.html}}, which have been linked
to nicotine dependence, based on the statistical significance of allele frequency difference between
cases and controls. These 288 genes are those for which mouse orthologs are found in the AGEA, with 
at least two datasets, one sagittal and one coronal.\\

The brainwide gene-expression data are used to compute co-expression networks.
 Restrictions to marker genes for brain regions defined by classical
 neuroanatomy \cite{AllenAtlas, markerGenesPaper}, or spatial clusters of gene-expression data \cite{ML, clusters}
 could be used in order to work out the anatomical properties of co-expression, in order to
 interpret them in terms of connections \cite{MBA, injections}.

\section{Model and gene-expression data}

\subsection{Data: the Allen Gene Expression Atlas (AGEA)}

The Allen Gene Expression Atlas is a high-resolution, brain-wide dataset,
that provides estimators of the number of mRNAs in the mouse brain at a resolution 
of 200 microns can be presented in the form of a voxel-by-gene matrix whose columns
represent genes, and whose lines corresponds to voxels ({emph{i.e}} cubes of side 200 microns into which the brain is partitioned):
\begin{equation}
E( v,g ) \simeq {\mathrm{Number\;of\; mRNAs\;for\; gene}} \;g \;{\mathrm{\;at\; voxel\;}}v.
\end{equation}
The entries of the matrix $E$ are 
gene expression energies were obtained from the co-registration to a reference
brain atlas of sets of ISH images
of thousands of genes in the Allen
Gene Expression Atlas \cite{AllenAtlasMol}. For each of the genes, an 
eight-week old C57Bl/6J male mouse brain was prepared as unfixed, 
fresh-frozen tissue.
 The following steps 
 were taken in an
automatized pipeline \footnote{For more details on the
processing of the ISH image series, see the NeuroBlast User Guide,
{\ttfamily{http://mouse.brain-map.org/documentation/index.html}}}:\\ 
\begin{itemize}
\item {\bf{Colorimetric {\it{in
        situ}} hybridization}};
\item {\bf{Automatic processing of the
    resulting images.}} In this step, tissue areas are found by  eliminating artefacts, looking 
for cell-shaped objects of size $\simeq 10-30$ microns;\\ 
\item {\bf{Aggregation of the raw pixel data into a grid.}}
The mouse brain is partitioned into $V = 49,742$ cubic voxels of
side 200 microns. All series of brain tissue 
are registered to a reference atlas \cite{AllenAtlas}.
 For each voxel $v$, the {\it{expression energy}} of
the gene $g$ is defined as a weighted sum of the greyscale-value
intensities of pixels $p$ intersecting the voxel:
\begin{equation}
E(v,g) := \frac{\sum_{p\in v} M( p ) I(p)}{\sum_{p\in v} 1},
\end{equation}
 where
$M( p )$ is a Boolean mask worked out by step 2 with value 1 if the
pixel is expressing and 0 if it is non-expressing.\\ 
\end{itemize}
The present analysis is focused on genes for which sagittal and coronal 
data are available. We computed the correlation coefficients
between sagittal and coronal data and selected the genes in the top-three 
quartiles of correlation ($G=3041$ genes), and used the coronal atlas \footnote{A searchable list of genes, consisting of all the genes from the AGEA, is available on-line as
part of the Brain Architecture project: {\ttfamily{http://addiction.brainarchitecture.org}}.
Heat maps of maximal-intensity projections of these genes, as well as visualization tools of their 
co-expression networks. 
}.

\subsection{Co-expression matrices}
Since each gene in the AGEA has a gene-expression energy 
given by a positive number at each of the $V$ voxels,
the columns of the matrix $E$ of gene-expression data are naturally identified to vectors in a $V$-dimensional space.
This is a very high-dimensional space. However, given two genes, the two corresponding columns of the 
matrix $E$ span a two-dimensional vector-space. The simplest geometric quantity
to study for this system of two vectors is the angle at which they intersect. As all the entries of the matrix $E$
are positive by construction, this angle is between 0 and $pi/2$. The angle between the two vectors
is  therefore completely characterized by its cosine, which is readily expressed in terms of the normalized
columns of the matrix of gene-expression energies.\\

Consider the matrix $\tilde{E}$ of $L^2$-normalized columns of $E$:
$$ \tilde{E}( v, g ) = \frac{1}{\sqrt{\sum_{v = 1}^VE(v,g ) ^ 2}} E(v,g ).$$
The $g$-th columns of $\tilde{E}$ is the direction of the expression-vector of gene $g$ (the factor 
$\sqrt{\sum_{v = 1}^VE(v,g ) ^ 2}$ is the Euclidean norm of the expression vector of gene $g$
in voxel space).\\

For any two genes $g$ and $g'$,
 consider the following scalar product:
\begin{equation}
{\mathrm{coExpr}}( g, g')= \sum_{v = 1}^V \tilde{E}( v, g )\tilde{E}( v, g' ).
\end{equation}
 This
number is between 0 and 1 by construction. It is the cosine of the
angle between the direction of the expression vectors of $g$ and
$g'$. We call it the {\emph{co-expression}}, or
{\emph{cosine similarity}}, of genes $g$ and $g'$.
The more co-expressed $g$ and $g'$ are in the brain, the closer their {\emph{cosine similarity}} is to 1.

These numbers, which we computed for pairs of genes in the AGEA
 are naturally arranged in a matrix, denoted by $C^{\mathrm{atlas}}$, with the genes 
arranged in the same order as the list of genes in the AGEA:
\begin{equation}
{C^{\mathrm{atlas}}}(g,g')= {\mathrm{coExpr}}( g, g'). 
\end{equation}
This matrix is symmetric because the scalar product is.
Its diagonal entries are all equal to one, as they correspond to the scalar product of normalized 
gene expression vectors with themselves. This diagonal is trivial in the sense that
it expresses that the vector of expression energies of each gene in the atlas is perfectly
aligned with itself. When we consider the distribution of the entries of the co-expression matrix, we really mean
the distribution of the upper-diagonal coefficients.\\

\subsection{Co-expression matrices as co-expression graphs}
Given a set of genes curated from the literature, possibly 
studied using different methods, one may ask if these genes (or a subset) expression
 profiles across the brain are particularly close to each other. 
The study of the co-expression matrix is a quantitative way to assess how exceptional 
a set of genes is.\\

  A set of strongly 
co-expressed genes corresponds to a submatrix
of the co-expression matrix {\emph{with large coefficients}}.
In order to formalise this idea,
we need to choose a way to define what a submatrix {\emph{with large coefficients}} is.
 We propose to study the matrix in terms of the underlying graph, because this
makes the results less dependent on the way genes are ordered. 
 Even though considering the data as a matrix suggested the study of the angles between its column vectors,
we are only interested in the set of co-expression between pairs of genes, and the set of pairs of genes 
does not depend on the order in which we present the genes of AGEA. \\

There are $G!$ ways 
of ordering the $G$ genes in the Allen Gene Expression Atlas. Generically they will generically give
rise to different co-expression matrices, related by similarity transformation. But the sets 
of highly co-expressed genes are certainly invariant under these operations.\\

The co-expression matrix can be mapped to a weighted graph in a straightforward way (see Figure (\ref{graphMotivation}) for a toy-model with 9 genes).\
 The vertices
of the graph are the genes, and the edges are as follows:\\
- genes $g$ and $g'$ are linked by an edge if their co-expression $C_{gg'}$ is strictly positive.\\
- If an edge exists, it has weight $C_{gg'}$.\\
 
Consider a set of genes of size $K$, for all of which data are available in the AGEA. 
 They correspond to indices $(g_1,\dots, g_K)$ in the columns of the matrix $E$ of gene-expression
data. We can construct 
a co-expression matrix just by extracting the coefficients of the co-expression matrix of the atlas 
corresponding to these genes. Let us denote this matrix $C^{\mathrm{set}}$:
\begin{equation}
C^{\mathrm{set}} = \left(C^{\mathrm{atlas}}(g,g') \right)_{g,g' \in \{g_1,\dots,g_K \}}.
\end{equation}
We would like to compare the properties of the matrix $C^{\mathrm{set}}$ to the ones 
of $C^{\mathrm{atlas}}$. Some quantities such as the average of the upper-diagonal coefficients
can be readily computed for both matrices, there is a sample-size bias that prevents us from comparing 
the results directly. Instead, we are going to study some properties of the graph underlying $C^{\mathrm{set}}$,
 and to compare them to the properties of the graphs underlying submatrices of $C^{\mathrm{atlas}}$ of the same size.

\subsection{Thresholding a co-expression graph}

We have not formalized the notion of a large co-expression coefficients.
 It is most likely impossible to give a useful absolute definition of properties
that would characterize large co-expression matrices, because the absolute values
of co-expression depend a lot on the resolution and will not be reproduced in other atlases 
and/or datasets. For instance, in the limit of a very coarse resolution, the 
data will not reflect much of the microscopic details of the gene-expression profiles,
and the entries of the co-expression matrix will be larger on average than in the present atlas. In the limit of a very 
fine resolution, for instance if one took the Allen data at full (one micron) resolution, the shape of the soma 
would become visible, and the expression is zero outside the soma. Even after co-registration
of all the ISH image series to the  Reference Atlas, the spatial distribution of somas would 
vary from brain to brain and the co-expression coefficients would therefore be much lower 
on average that at a resolution of 200 microns. \\

\begin{figure}
\begin{center}
\begin{minipage}{0.48\textwidth}
\includegraphics[width=2.25in,keepaspectratio]{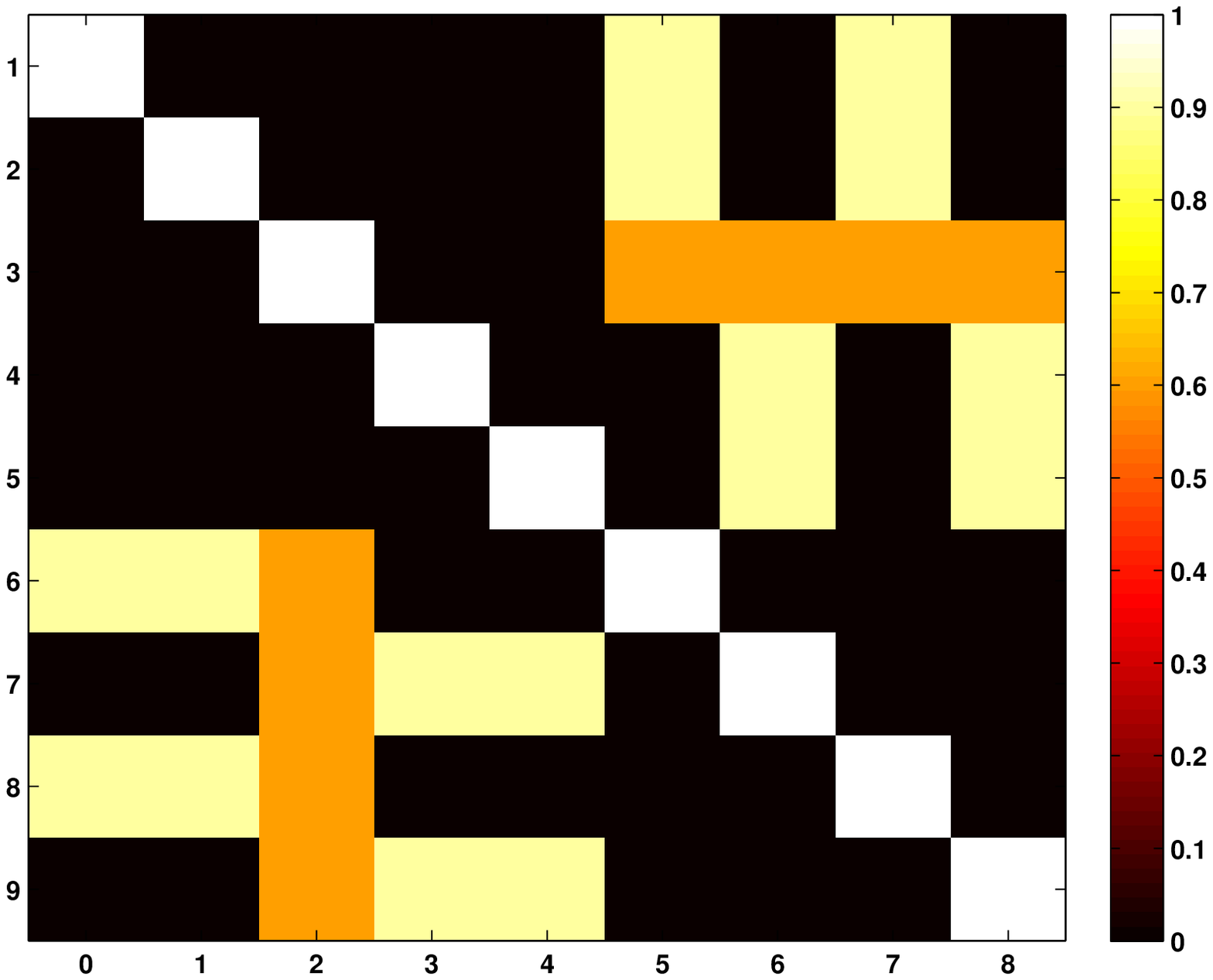}
\end{minipage}
\begin{minipage}{0.48\textwidth}
\includegraphics[width=2.25in,keepaspectratio]{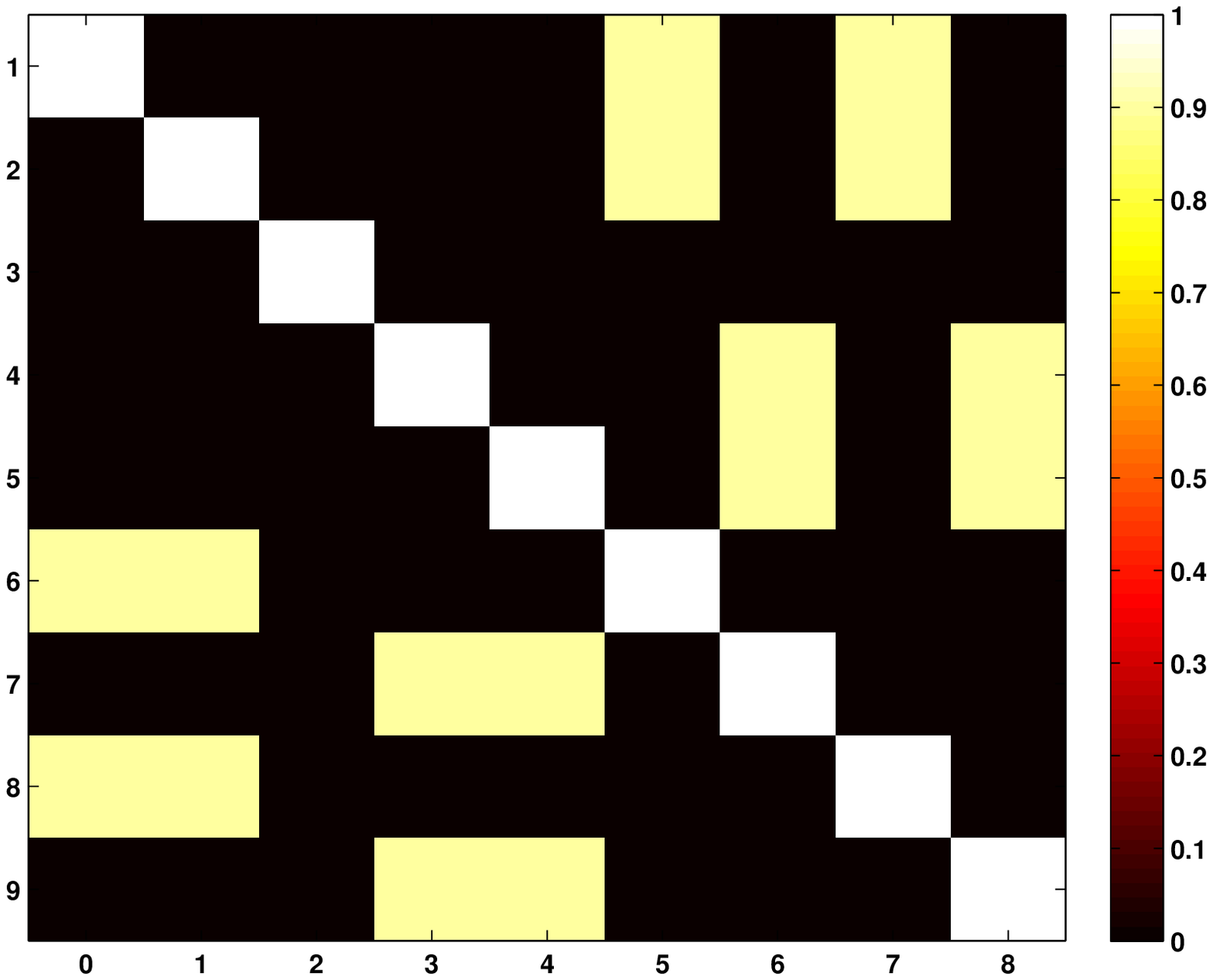}
\end{minipage}

\begin{tikzpicture}[node distance=1.3cm,>=stealth',bend angle=45,auto]

  \tikzstyle{place}=[circle,thick,draw=blue!75,fill=blue!20,minimum size=6mm]
  \tikzstyle{red place}=[place,draw=red!75,fill=red!20]
  \tikzstyle{transition}=[rectangle,thick,draw=black!75,
  			  fill=black!20,minimum size=4mm]

  \tikzstyle{every label}=[red]

  \begin{scope}
    \node [place] (g1)                                    {$g_1$};
    \node [place] (g2) [below of=g1]                      {$g_2$};
    \node [place] (g3)  [below of= g2] {$g_3$};
    \node [place] (g4) [below of=g3]                       {$g_4$};
    \node [place] (g5) [below of=g4]                      {$g_5$};

    \node [place] (g6) [left of=g2] {$g_6$}
      edge [bend left, color=red]            (g1)
      edge [bend right, color=blue]               (g3)
      edge [color=red]             (g2);

    \node [place] (g7) [left of=g4] {$g_7$}
      edge [bend right, color=red]                 (g5)
      edge [bend left, color=blue]                 (g3)
      edge [color=red]               (g4);

    \node [place] (g8) [right of=g2] {$g_8$}
      edge [color=red]                   (g2)
      edge [bend left, color=blue]  node{0.6}                 (g3)
      edge [bend right, color=red] node[swap] {0.9} (g1);

    \node [place] (g9) [right of=g4] {$g_9$}
      edge [color=red]                            (g4)
      edge [bend right, color=blue]           (g3)
      edge [bend left, color=red]  node{0.9}       (g5);
  \end{scope}

  \begin{scope}[xshift=6.5cm]
    \node [place]
                      (g1')                                                {$g_1$};
    \node [place]     (c1') [below of=g1']                                 {$g_2$};
   \node [place] (s1') [below of=c1']      {$g_3$};
   
    \node [place]     (c2') [below of=s1']                      {$g_4$};
    \node [place]
                      (w2') [below of=c2']                                 {$g_5$};

    \node [place] (e1') [left of=c1'] {$g_6$}
      edge [bend left, color=red]                  (g1')
      edge [color=red]                           (c1');

    \node [place] (e2') [left of=c2'] {$g_7$}
      edge [bend right,color=red]                 (w2')
      edge [color=red]                           (c2');

    \node [place] (l1') [right of=c1'] {$g_8$}
      edge [color=red]                            (c1')
      edge [bend right, color=red] node[swap] {0.9} (g1');

    \node [place] (l2') [right of=c2'] {$g_9$}
      edge [color=red]                            (c2')
      edge [bend left,color=red]  node {0.9}       (w2');
  \end{scope}

  \draw [-to,thick,snake=snake,segment amplitude=.4mm,
         segment length=2mm,line after snake=1mm]
    ([xshift=5mm]g3 -| g8) -- ([xshift=-5mm]s1' -| e1')
    node [above=1mm,midway,text width=3cm,text centered]
      {Thresholding of the co-expression};

  \begin{pgfonlayer}{background}
    \filldraw [line width=4mm,join=round,black!10]
      (g1.north  -| g8.east)  rectangle (g5.south  -| g6.west)
      (g1'.north -| l1'.east) rectangle (w2'.south -| e1'.west);
  \end{pgfonlayer}
\end{tikzpicture}
\end{center}
\caption{A toy model with 9 genes, and only three distinct values of co-expression, 0, 0.6 and 0.9, for simplicity.
 The two heat maps of the co-expression matrix correspond 
to two different orderings of the genes, but the co-expression coefficients are the same. On the second heat map,
it is obvious by eye that the first four genes form an 'island' with high co-expression, but it is hard 
to observe this on the first map, simply because these genes are not placed at adjacent columns. However, plots of the underlying graph show that these genes stay connected even when the links with strength weaker than a
threshold are pruned.}
\label{graphMotivation}
\end{figure}

 Hence we have to define large co-expression matrices in relative terms,
using thresholds on the value of co-expression that describe 
the whole set of possible values.
The coefficients of the co-expression networks are numbers between $0$
and $1$ by construction. We define the following thresholding procedure on co-expression
graphs: given a threshold $\rho$ between $0$
and $1$, and a co-expression matrix denoted by $C$ (which can come from any 
set of genes in the AGEA), put to zero all the
coefficients that are lower than this
coefficient. This graph comes from a thresholded co-expression matrix $C_\rho$
such that the underlying graph has only edges with weight larger than the threshold:
\begin{equation}
C_\rho(i,j) = C(i,j) \times\mathbf{1}( C(i,j) \geq \rho).
\end{equation}
By construction we have $C_0 = C$. The graph corresponding to the matrix $C_\rho$ has connected components, 
and each connected component has a certain number of genes in it. Our notion 
of a large co-expression matrix, relative to a threshold $\rho$ is therefore a topological property of the graph (see Figure \ref{graphMotivation} for an illustration of a toy-model
with only 10 genes) underlying
 the thresholded matrix $C_\rho$. 

If the initial co-expression matrix $C$  has size $G$ ({\emph{i.e.}} corresponds to $g$ genes), then for every 
integer $k$ between 1 and $G$ we can count the number 
$N_\rho(k)$ of connected components of $C_\rho$ that have exactly  $k$ genes in them.
 
We can study the average size of connected components
of thresholded co-expression networks
\begin{equation}
\mathcal{G}( \rho ) = \frac{\sum_{k = 1}^G k  N_\rho(k)}{\sum_{k = 1}^G  N_\rho(k) },
\end{equation}
and the size of the largest connected component:
\begin{equation}
\mathcal{A}( \rho ) = \frac{\sum_{k = 1}^G k  N_\rho(k)}{\sum_{k = 1}^G  N_\rho(k) },
\end{equation}
\begin{equation}
\mathcal{M}( \rho ) = \mathrm{max}\left\{k \in [1..G],  N_\rho(k) > 0 \right\},
\end{equation}
as a function of the threshold $\rho$. We can see that 
$\mathcal{A}( 0 )$ 
is the size of the set of genes, as the whole set is connected. At large thresholds every single singe is disconnected from the 
other genes, as having co-expression equal to one is equivalent to having exactly the same expression across the whole brain. 
So at threshold 1 all the connected components have size one, and $\mathcal{G}( 1 ) = 1$. Examples are given below for 
several values of $G$.\\

\subsection{Monte Carlo study}
In order to eliminate the sample-size bias we have to compare  
graph properties of $C^{\mathrm{set}}$ to the graph properties of matrice sof the same size, drawn from $C^{\mathrm{atlas}}$. 
If $C^{\mathrm{set}}$ has size $K$, we can  repeatedly draw random sets
of genes of size $K$, extract the corresponding submatrices of the 
full co-expression matrix $C^{\mathrm{atlas}}$, and compute the 
size of the largest connected component for each value of the threshold for which $C^{\mathrm{set}}$ was studied.\\

In order to explore the space of thresholds we have to choose a discrete set of threshold regularly 
spaced between 0 and 1. Call $R$ the number of random draws. The computations can be described as follows
 in pseudocode:\\
{\ttfamily{
1. Choose a number of thresholds $T$ to study.\\
2. Choose a number of draws $R$ to be performed for each value\\ of
the threshold.\\
3. For each integer $t$ between $1$ and $T$:\\
3.a. consider the threshold $\rho_t = \frac{t}{T};$\\
3.b. compute the connected components of the thresholded matrix $C^{\mathrm{set}}_{\rho_t}$;\\ 
call $\mathcal{M}^{\mathrm{set}}( \rho )$ the size of the largest connected component;\\
4. for each integer $r$ between $1$ and $R$:\\
draw a random set of distinct indices of size $K$ from [1..G],\\
 extract the corresponding submatrix
of $C^{\mathrm{atlas}}$;\\ call it $C^{r}$, and repeat step 3 after substituting  $C^{r}$ to  $C^{set}$;\\
call  $\mathcal{M}^{\mathrm{r}}( \rho_t )$ the size of the largest connected component 
 of $C^{\mathrm{r}}_{\rho_t}$.
 }}
The two values chosen in the first step are obviously dictated by the speed on the computation. 
The number of mathematical operation is $O(TR)$. If interesting and/or regular behaviour is observed
around a particular threshold, the algorithm can be rerun on a set of threshold centered around these values,
 because the results of loops associated to different values of the threshold are independent.\\

At each value $\rho$ of the threshold, we therefore have:\\
- the size of the maximal connected component $\mathcal{M}^{\mathrm{set}}( \rho )$,\\ 
- a distribution 
of $R$ numbers, each of wchich is the size of the largest connected components of 
a random submatrix of the same size as the set of genes to study, thresholded at $\rho$.\\
We can study where in $\mathcal{M}^{\mathrm{set}}( \rho )$ sits in the distribution
by computing the number of standard deviations by which it deviates from the mean across all draws:
$$\mu( \rho) = \frac{1}{R} \sum_{r = 1}^R \mathcal{M}^{\mathrm{r}}( \rho ),$$
$$\sigma(\rho) := \sqrt{ \frac{1}{R}\sum_{r = 1}^R ( \mathcal{M}^{\mathrm{r}}( \rho ) -\mu(\rho) )^2},$$
$$\delta( \rho ):= \frac{\mathcal{M}^{\mathrm{set}}( \rho )}{\sigma(\rho)}$$

\subsection{Empirical cumulative distribution functions of co-expression coefficients}
\subsubsection{Empirical distribution function}
In order to complement the graph-theoretic approach, we can study the cumulative distribution
function of the co-expression coefficients in the special set, and compare it 
to the one resulting from randonm sets of genes of the same size.

Let us plot the empirical distribution functions of the coefficients above the
diagonal in the co-expression matrices $C$, for $C = C^{\mathrm{atlas}}$ in blue, 
for $C = C^{\mathrm{set}}$ in red. These distribution functions are evaluated in the following way:
Let $N$ denote the size of the matrix $C$, i.e. the number of genes from which $C$ was computed. Consider
the coefficients above the diagonal (which are the meaningful quantities in $C$ by construction) and arrange them into a
vector $C_{\mathrm{vec}}$ with $N' = N(N-1)/2$ components:
$C_{{\mathrm{vec}}} = \{ C_{gh} \}_{1 \leq g \leq N, h > g}$. The components are numbers between 0 and 1. For every number between
0 and 1, the empirical distribution function of  $C$, denoted by ${\mathrm{edf}}^C$ is defined as the fraction of the components of 
$C_{{\mathrm{vec}}}$ that are smaller than this number:
$$ {\mathrm{edf}}^C : [ 0,1 ] \rightarrow [ 0,1 ]$$
$$ x \mapsto \frac{1}{N'} \sum_{k = 1}^{N'} \delta_{
  C_{{\mathrm{vec}}}( k ) \leq x }.$$

\subsubsection{Bootstrapping: cumulative distribution functions of random sets of genes} 
We would like  to compare the network of interest to random networks of the same size, drawn from the 
set of genes in our dataset. The procedure is exactly the same as with the thresholded matrices, except that 
the quantities computed from the random draws are cumulative dictribution functions rather than connected components.
Let us repeteadly draw random subsets of genes with the same number of elements as the set of interest, compute the
empirical distribution function of the corresponding submatrix of $C^{\mathcal{T}}$ and average over the 
draws. The average over the draws should converge towards the  one of a typical
network of $N$ genes, when the number of draws becomes larger.\\

\section{Application to a list of addiction-related genes}

\subsection{A set of 288 genes from NicSNP}

\begin{figure}
\center
\includegraphics[width=4.5in,keepaspectratio]{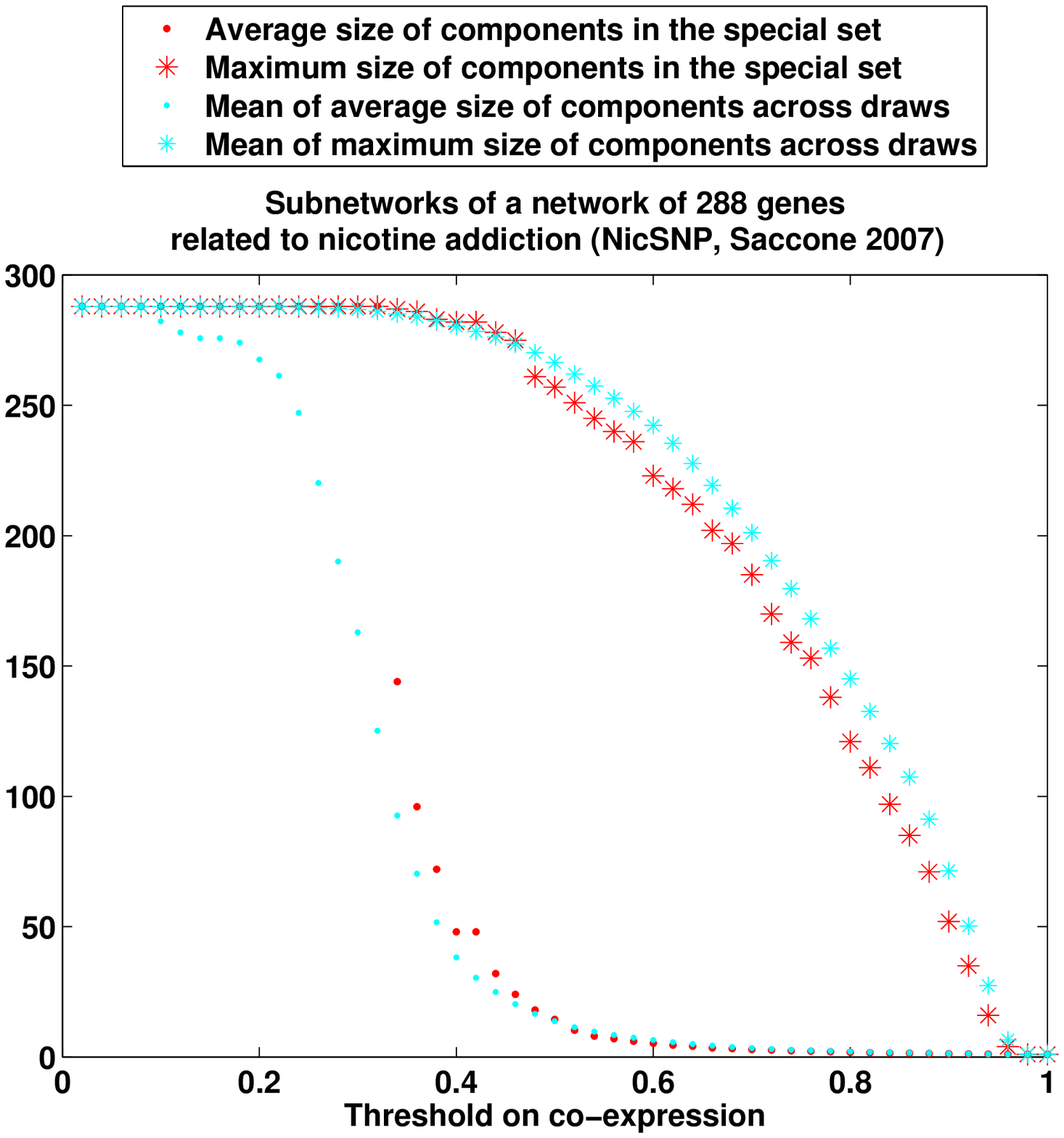}\\
\includegraphics[width=4.5in,keepaspectratio]{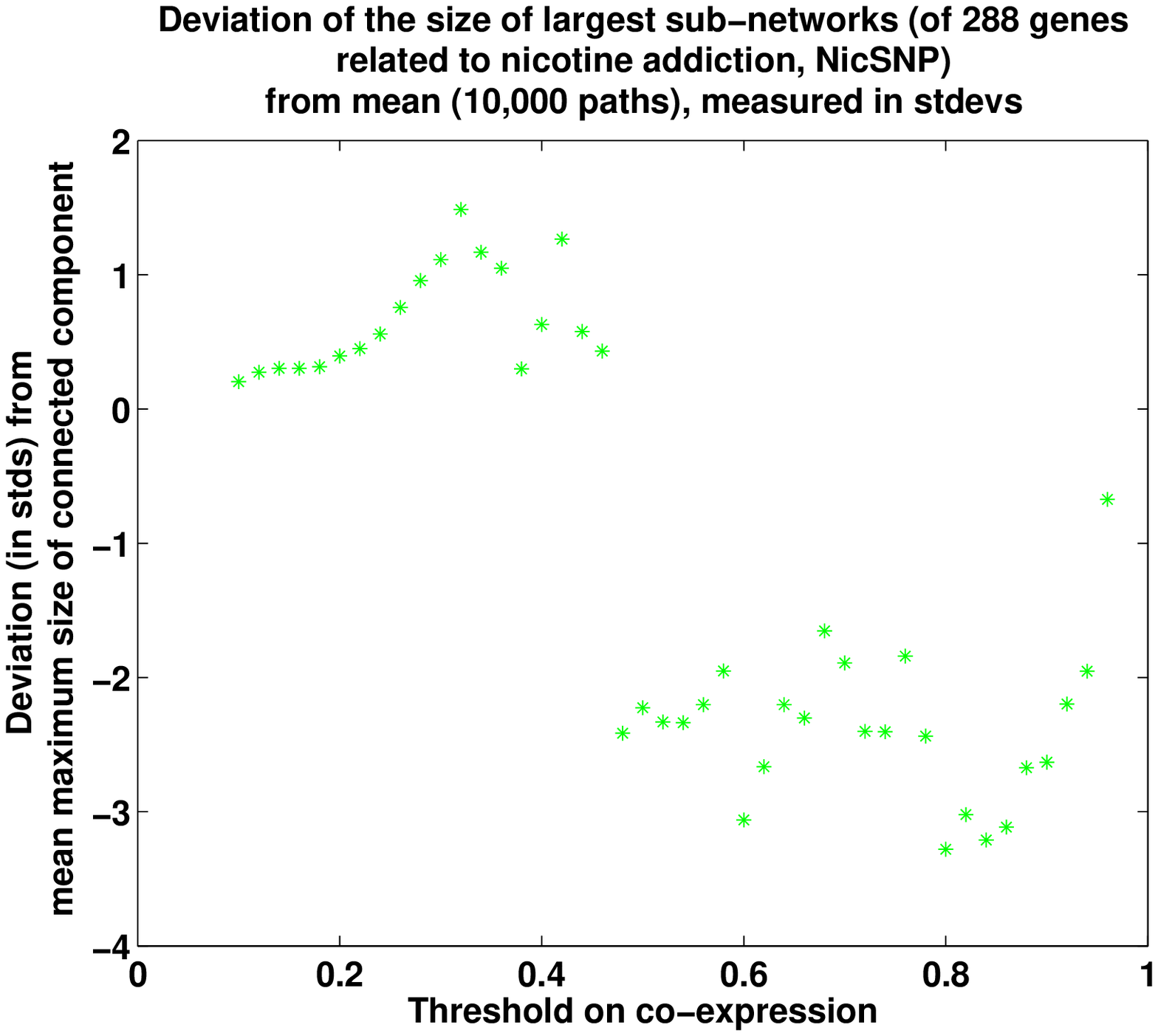}\\
\caption{Statistics of maximum and average size of connected components
of thresholded co-expression graphs (addiction-related genes in red, random sets in cyan), 288 genes.}
\label{fig:nicotineAll}
\end{figure}

We applied the graph-theoretic method described above to a set of 288 genes
related to nicotine addiction gathered from the NicSNP database \cite{Saccone}.
The results are presented on Figure (\ref{fig:nicotineAll}). The special set of 288 genes 
is not more co-expressed than expected by chance, since thresholding at high values of the 
co-expression coefficient induces connected components that are smaller on average than those of random 
sets of genes of size 288.
 The  result is therefore rather negative : there is no clear statistical significance of the connected 
component of the co-expression matrix of these genes, especially at the highest levels of co-expression.
  The use of the proposed graph-theoretic method is however adapted to the determination of  exceptional sets 
of genes that are small in scale of a pre-determined set of condition-related genes.\\

\begin{figure}
\center
\includegraphics[width=4.5in,keepaspectratio]{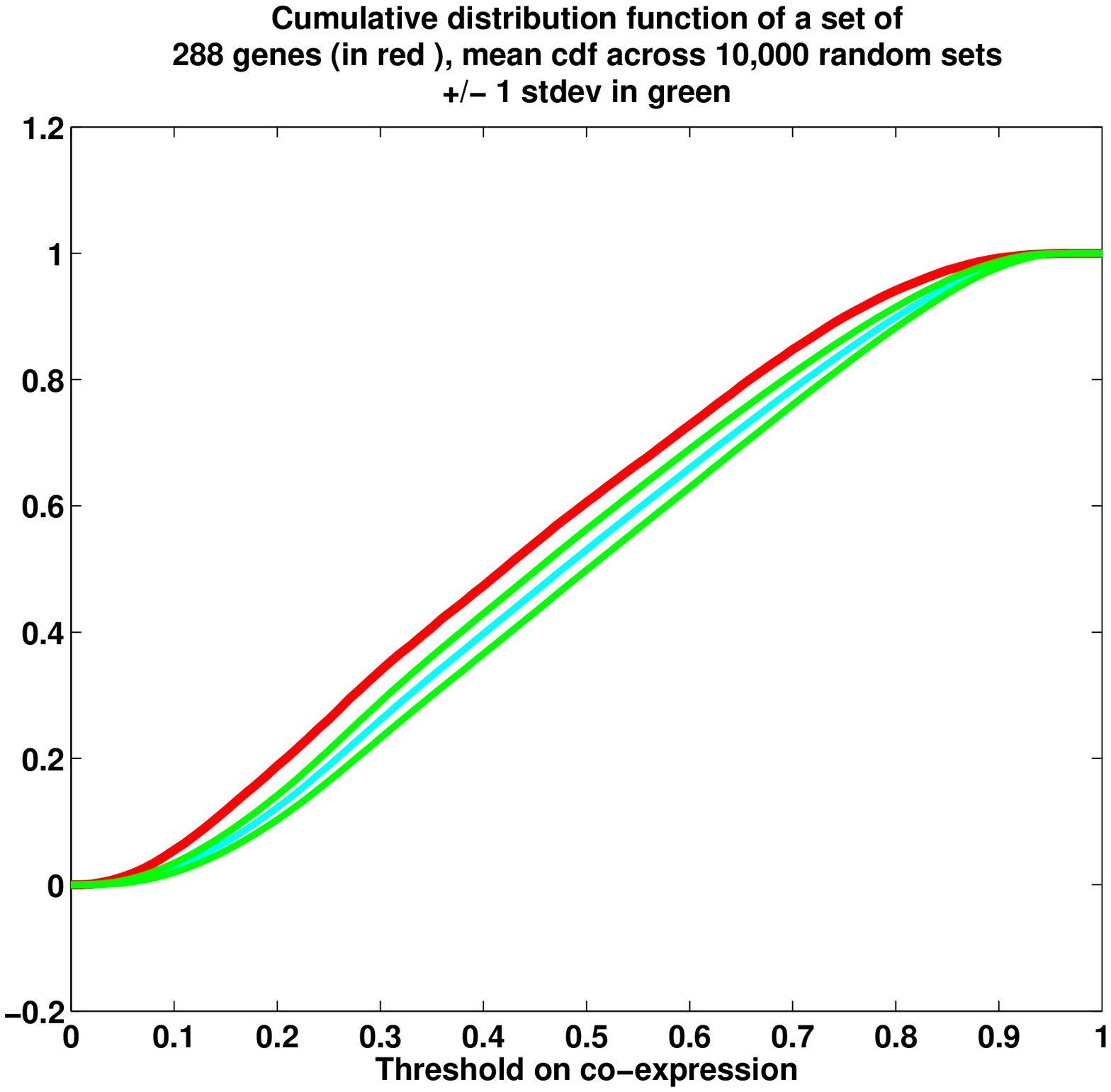}
\caption{Cumulative distribution function of cosine similarities in co-expression graphs of 288 genes related to nictine addiction.}
\label{fig:Nic288Cumul}
\end{figure}

The study of the cumulative distribution function of co-expression for the full set of 288 genes
is also rather negative (see Figure \ref{fig:Nic288Cumul}), as the cdf grows faster than expected by chance for small values of 
the co-expression.

\subsection{Search for statistically significants subsets of fixed size}

It may be the case that subsets of the list of genes to study exhibit 
exceptional co-expression properties, while the whole set is not particularly co-expressed.  
In order to prioritize subsets of the genes in NicSNP for further study, we would like to identify 
 such subsets.\\

We would be in that case for instance if the first few genes had all very-high co-expression with each other, while the next 
genes are orthogonal to them as well as with each other {\footnote{This very extreme and idealized  case is geometrically possible within the AGEA
at a resolution of 200 microns, because
the number of genes in the dataset is smaller than the number $V=49,742$  of voxels. The gene-expression vectors corresponding to
all the columns of the data matrix can all be non-zero and still be orthogonal because they are elements of  a $V$-dimensional space.}}, corresponding to a co-expression matrix
with a square of high coefficients (of size, say, $G_{high}$) in the upper-left corner, and zeros everywhere. 
However, the order in which genes are presented does not ensure that the set
of highly expressed genes should be the set of the first $G_{high}$ genes.\\

In order to detect such subsets of genes, one can allow the user to specify a list $I_{init}$ few indices in the list 
of genes, that have been observed to have high co-expression, or that are of special interest as a subset.

For a thorough search of exceptional sets of co-expressed genes of a given size, one can take the list $I^{init}$ to consist of just one index, and repeat the procedure for 
all posible choice of this index.\\

Let the list $I^{init}$ consist of  $G_{init}$ indices:
$$I^{init}=: ( i_1, \dots, i_{G_{high}} ).$$
One can grow it one element at a time by adding to it the gene that co-expresses
the most on average with the genes that are already in the list. That is, for each
gene whose index is not in $I^{init}$, compute the sums of co-expressions with genes
whose  index is in $I^{init}$, and pick the index $g_{suppl}$ that maximizes the sum:
$$c^{extra}:=\left( \sum_{j\in I^{init}}C^{small}_{ij} \right)_{i\notin I_{init}}.$$
$$g_{suppl} := {\mathrm{arg max}}_{i\notin I^{init}} \left(c^{extra}_i \right).$$

The gene with index $g_{suppl}$ is then added to the list:
$$I^{init}\rightarrow [ I^{init},g_{suppl} ],$$
and the co-expression
matrix corresponding to the new list can be studied by the thresholding technique described 
above.
\begin{figure}
\center
\includegraphics[width=4.5in,keepaspectratio]{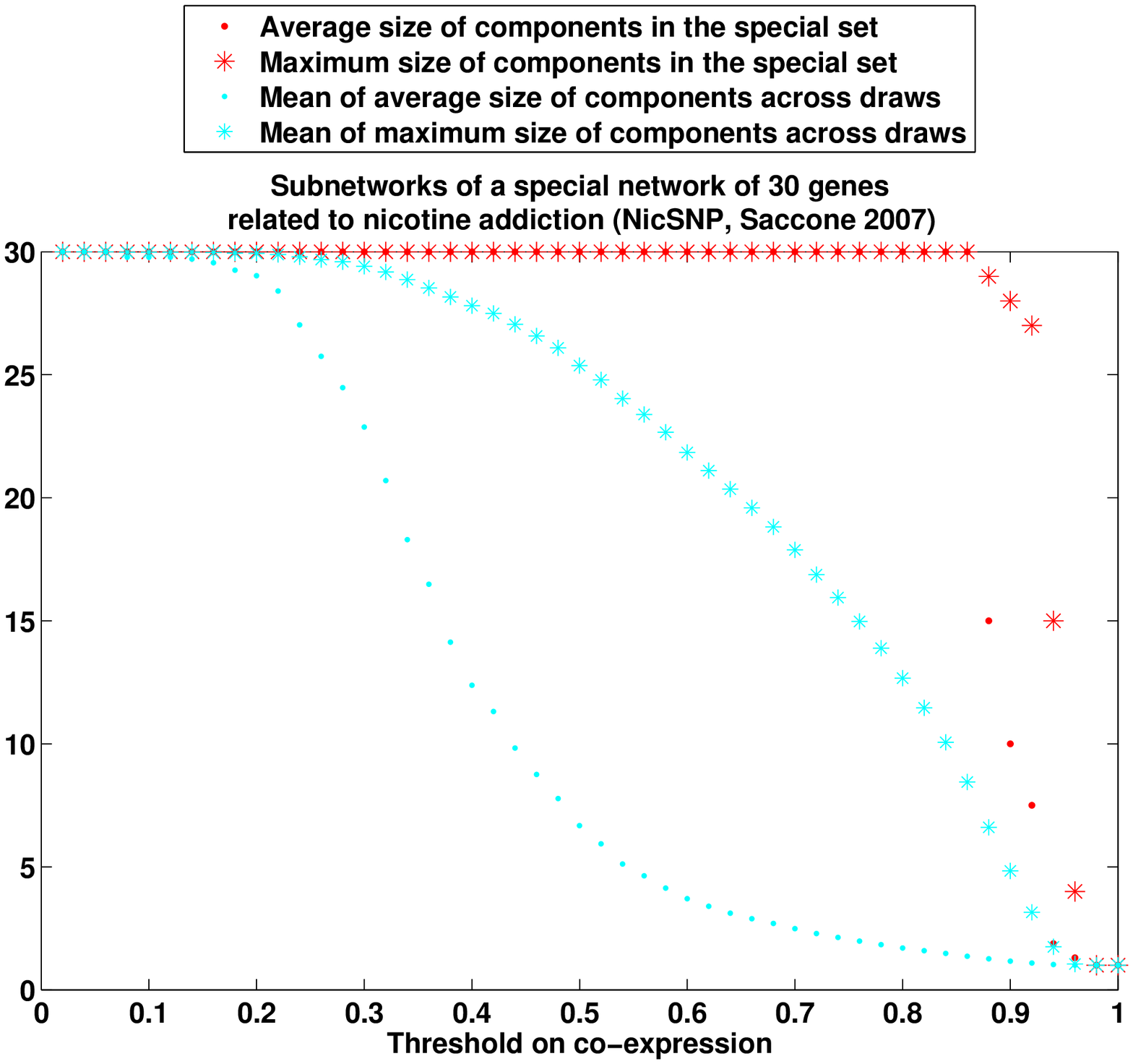}\\
\includegraphics[width=4.5in,keepaspectratio]{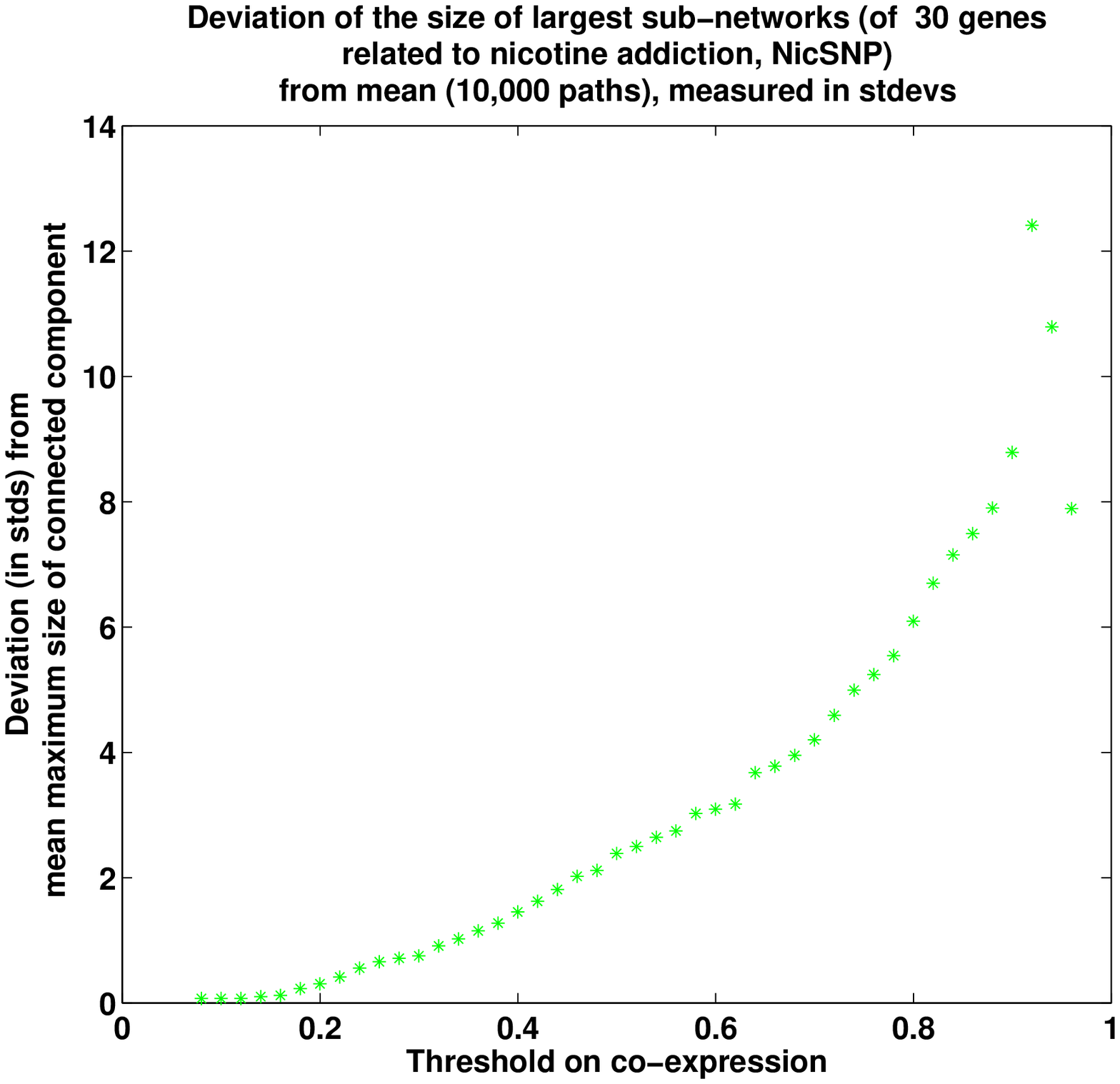}\\
\caption{Statistics of maximum and average size of connected components
of thresholded co-expression graphs (addiction-related genes in red, random sets in cyan), 30 genes.}
\label{fig:addictioncoExprIslands30}
\end{figure}
This procedure was used to construct a set of 30 genes, which was studied 
using the graph-theoretic procedure. The names of these genes are the following:
 {\emph{Gprasp1, Uchl1, Grin1, Ctsb, Gria2, Phyh, Snap25, Actr2, Gabarapl1, Calm1, Dlgh4, Syt1, Ppp1r9b, Cttn, Grik5, Gria3, 
Slc1a2, Ssbp4, Gria4, Hint1, Atrx, Per1, Slc1a1, Gabbr2, Chrm1, Gtpbp9, Cap1, Fbxw2, Mtch2, Socs5}}.\\

The results are shown of Figure (\ref{fig:addictioncoExprIslands30}),
from which it is quite clear that this set of genes is exceptionally co-expressed.
The procedure can then be repeated on the set of 258 nicotine-related genes obtained by removing those genes.\\

\begin{figure}
\center
\includegraphics[width=4.5in,keepaspectratio]{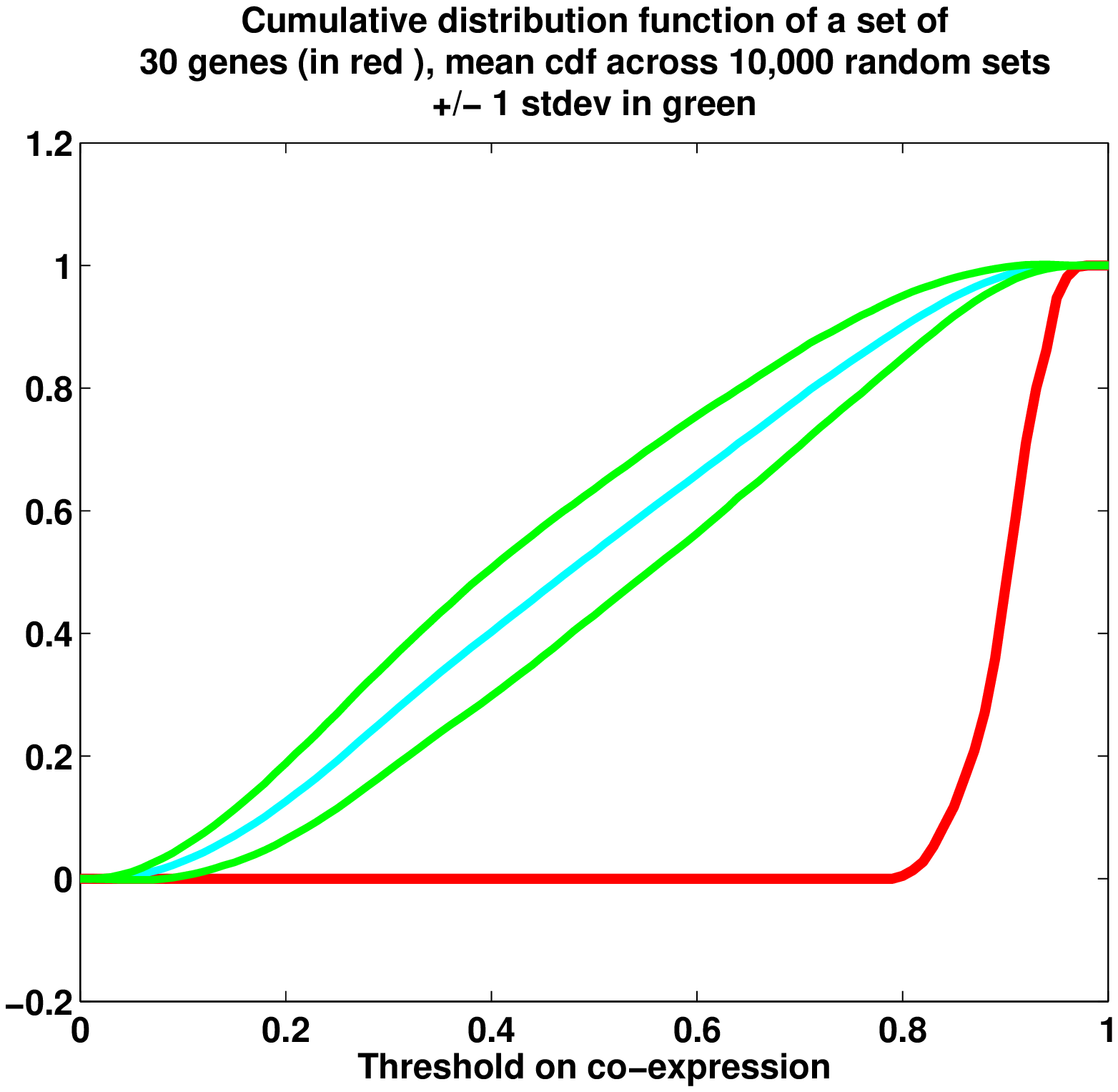}
\caption{Cumulative distribution function of cosine similarities in co-expression graphs of 30 genes related to nicotine addiction (in red).}
\label{fig:Nic30Cumul}
\end{figure}

The study of the cumulative distribution function of co-expression for this special set of 30 genes
 deviates  (see Figure \ref{fig:Nic30Cumul}), as the cdf takes off at higher values of 
the co-expression than expected by chance.

\section*{Acknowledgments}
 This research is support by the
NIH/NIDA grant 1R21DA027644-01.\\

\end{document}